\documentclass[10pt, conference, compsocconf]{IEEEtran}

\usepackage{ifpdf}
\ifCLASSINFOpdf
  \usepackage[pdftex]{graphicx}
\else
  \usepackage[dvips]{graphicx}
\fi

\usepackage{amsmath, amsthm, amssymb}
\newtheorem{definition}{Definition}

\begin{document}
\title{Significance Relations for the Benchmarking of Meta-Heuristic Algorithms}


\author{\IEEEauthorblockN{Mario K\"oppen and Kei Ohnishi}
\IEEEauthorblockA{Computer Science and Engineering Department\\
Kyushu Institute of Technology\\
680-4 Kawazu, Iizuka, 820-8502 Fukuoka, Japan\\
Email: mkoeppen,ohnishi@cse.kyutech.ac.jp}}

\maketitle

\begin{abstract}
The experimental analysis of meta-heuristic algorithm performance is usually based on comparing average performance metric values over a set of algorithm instances. When algorithms getting tight in performance gains, the additional consideration of  significance of a metric improvement comes into play. However, from this moment the comparison changes from an absolute to a relative mode. Here the implications of this paradigm shift are investigated. Significance relations are formally established. Based on this, a trade-off between increasing cycle-freeness of the relation and small maximum sets can be identified, allowing for the selection of a proper significance level and resulting ranking of a set of algorithms. The procedure is exemplified on the CEC'05 benchmark of real parameter single objective optimization problems. The significance relation here is based on awarding ranking points for relative performance gains, similar to the Borda count voting method or the Wilcoxon signed rank test. In the particular CEC'05 case, five ranks for algorithm performance can be clearly identified.
\end{abstract}


% \fi
\IEEEpeerreviewmaketitle

\section{Introduction}
Recently there is growing interest in the experimental analysis of algorithm performance. The establishment of computational paradigms like soft computing and computational intelligence has lead to a rapidly increasing number of new algorithm proposals, esp. based on computational models of evolution, genetics, or swarms intelligence, but also modifications of hitherto uniform algorithms to the level of the appearance of new algorithms, or combination, fusion and hybridization of existing algorithms into new ones. The common aspects of these algorithms --- often called meta-heuristic algorithms --- is that their design is essentially problem-independent, that their processing usually includes random factors, and that there are no guarantees or known bounds for their performance while they posses parameters influencing the likelihood of good results by adjustable effort like population size or number of generations. Thus, the algorithm performance can differ from problem to problem, and on algorithm instance to algorithm instance, allowing for a performance competition between all those algorithms.

However, the experimental evaluation of algorithm performance faces a number of problems. Just to name the essential ones: (1) the problem of specifying a subset of test functions that are challenging enough to generate a spectrum of performance values, while avoiding any ``needle-in-a-haystack'' pure-chance search that would not provide any meaningful insights into strength and weaknesses of the studied algorithm; (2) fairness of comparison, usually understood as measuring performance under ``equal effort'' conditions like same number of test function evaluations (but commonly not memory usage or CPU time); (3) the means of quantifying the experimental results into performance measures and the way of comparing them (where the No-Free-Lunch theorems \cite{wolpert1997no} state the fundamental non-existence of a distinguishing measure for all possible functions but also \cite{koppen2004no} stating the non-existence of related benchmarks); (4) the question about favorable parameter settings and the related identity of an algorithm, which often allows for a number of structural modifications to be applied while the algorithm is still considered the same, and which of these design choices to tolerate while still maintaining a fair comparison.

A number of approaches tried to accommodate these problems and proposed a set of benchmark functions and related bounds on effort-related algorithm parameters. Among them, the series of benchmark suites presented at the annual IEEE Conference on Evolutionary Computation (CEC) gained a lot of attention. From 2005, where a general benchmark on evolutionary real parameter single objective optimization was presented as an open contests (as well as in 2010 and 2013), a number of benchmarks on various specific aspects followed: for example on evolutionary constrained real parameter single objective optimization problems in 2006, on large-scale single objective global optimization with bound constraints in 2008 and 2010, on niching in 2010. In between, the CEC'05 benchmark has become a de facto standard for the evaluation of new algorithms.

Subsequent publications like \cite{garcia2009study} showed that despite the well-thought and modern design of these benchmarks, the problem of a proper evaluation of the results remained an issue. This is also related to the growing acceptance of statistical methods in the evaluation of experimental algorithm performance, assuming that each algorithm unveils a statistical distribution of its performance values. The general consideration here is about significance of a performance improvement. While it has become common to repeat experiments 10 or 30 times and consider the numerical average of closest approaches to known extrema of a specific benchmark function as a suitable quantity, it became also clear that an algorithm, showing here performance $p + \epsilon$ is not automatically better than an algorithm with performance $p$ --- it has to be \emph{significantly} better as well. The meaning of significance then is usually related to a statistical confidence value that the average performances stemmed from different distributions (more precisely: to reject the $H_0$ hypothesis that for two algorithms both performance outcomes follow the same distribution). However, the notion of significance also introduces another relevant aspect in the question for the ``best algorithm'' that has not been solicited so far: the loss of an absolute comparison.

To illustrate the meaning of this, compare the situation with typical ways of performing sport contests, where we can easily identify two main lines. For some sports, we are considering absolute means of success. In 100-meter dash, the performance is just the time needed to pass the 100 meter distance. This is an \emph{absolute} value. The record performance can be saved and is available at any later time to decide on the setting of a new record performance. Currently, it is 9.58s, which is a result from 2009. In comparison, team sports like soccer, baseball, tennis or handball do not have such an absolute value of performance and is therefore performed in tournaments. It means from one match there is no quantity derived that allows to judge the performance of the next match (we are not considering global achievements that go into the ``hall of fame'' here). The performance evaluation is relative. This has a number of implications: for example, we can tolerate for intransitivity in a tournament, which would be nonsense in a race --- when team $A$ wins against team $B$ and team $B$ afterwards against team $C$ there is no reason to assume that in a later match team $A$ is guaranteed to win against team $C$. However, there are various attempts to assign ratings to teams that allow for the approximate computation (or prediction) of ranks and match outcomes, as for example Massey's approach of assigning ratings such that their difference is as close as possible to known match outcomes \cite{massey1997statistical}, or Keener's approach of rating inspired by the famous page rank procedure as it is used by Google(TM) search engine \cite{keener1993perron}.

Algorithm performance shares aspects with both kinds of sport contest evaluations: an algorithm does not need the parallel processing of another algorithm to assess its performance for some benchmark function (otherwise the study of algorithm performance would enter the realms of game theory). But as soon as we introduce the aspect of significance, there is no absolute single value measure available anymore and the justification of algorithm betterness becomes a pairwise exercise, thus a relative one.

Here we want to study a way to account for these aspects based on a purely (set) relational framework. We contribute a formal definition of a performance comparison with significance and identify a subset of relations where an optimal choice of the significance level is feasible, based on the relation between minimal number of cycles appearing in the comparison and the smallest number of best algorithms (section II). In section III we apply this framework to some results presented for the CEC'05 benchmark in order to demonstrate the design of feasible relations for comparison, followed by a discussion in section IV.

\begin{table*}
\renewcommand{\arraystretch}{1.3}
\setlength{\tabcolsep}{4pt}
\centering
\caption{The average performances for 100 runs on the 25 test problems of the CEC'05 benchmark in dimension 10 for 11 test algorithms.The values here are the same as the values used in \cite{garcia2009study}, with some adjustment of numerical scale for better readability.}
\label{tab:1}
\begin{tabular}{|c|ccccccccccc|}
\hline
 & BLX-GL50 & BLX-MA & CoEVO & DE & DMS-L-PSO & EDA & G-CMA-ES & K-PCX & L-CMA-ES & L-SaDE & SPC-PNX\\\hline
f1 & 1.00E-09 & 1.00E-09 & 1.00E-09 & 1.00E-09 & 1.00E-09 & 1.00E-09 & 1.00E-09 & 1.00E-09 & 1.00E-09 & 1.00E-09 & 1.00E-09\\
f2 & 1.00E-09 & 1.00E-09 & 1.00E-09 & 1.00E-09 & 1.00E-09 & 1.00E-09 & 1.00E-09 & 1.00E-09 & 1.00E-09 & 1.00E-09 & 1.00E-09\\
f3 & 5.71E+02 & 4.77E+04 & 1.00E-09 & 1.94E-06 & 1.00E-09 & 2.12E+01 & 1.00E-09 & 4.15E-01 & 1.00E-09 & 1.67E-02 & 1.08E+05\\
f4 & 1.00E-09 & 2.00E-08 & 1.00E-09 & 1.00E-09 & 1.89E-03 & 1.00E-09 & 1.00E-09 & 7.94E-07 & 1.76E+06 & 1.42E-05 & 1.00E-09\\
f5 & 1.00E-09 & 2.12E-02 & 2.133 & 1.00E-09 & 1.14E-06 & 1.00E-09 & 1.00E-09 & 4.85E+01 & 1.00E-09 & 0.012 & 1.00E-09\\\hline
f6 & 1.00E-09 & 1.49 & 1.25E+01 & 1.59E-01 & 6.89E-08 & 4.18E-02 & 1.00E-09 & 4.78E-01 & 1.00E-09 & 1.20E-08 & 1.89E+01\\
f7 & 1.17E-02 & 1.97E-01 & 3.71E-02 & 1.46E-01 & 4.52E-02 & 4.21E-01 & 1.00E-09 & 2.31E-01 & 1.00E-09 & 0.02 & 8.26E-02\\
f8 & 20.35 & 20.19 & 20.27 & 20.4 & 20 & 20.34 & 20 & 20 & 20 & 20 & 20.99\\
f9 & 1.154 & 0.4379 & 19.19 & 0.955 & 1.00E-09 & 5.418 & 0.239 & 0.119 & 44.9 & 1.00E-09 & 4.02\\
f10 & 4.975 & 5.643 & 26.77 & 12.5 & 3.622 & 5.289 & 7.96E-02 & 0.239 & 40.8 & 4.969 & 7.304\\
f11 & 2.334 & 4.557 & 9.029 & 0.847 & 4.623 & 3.944 & 0.934 & 6.65 & 3.65 & 4.891 & 1.91\\
f12 & 406.9 & 74.3 & 604.6 & 31.7 & 2.4001 & 442.3 & 29.3 & 149 & 209 & 4.50E-07 & 259.5\\
f13 & 0.7498 & 0.7736 & 1.137 & 0.977 & 0.3689 & 1.841 & 0.696 & 0.653 & 0.494 & 0.22 & 0.8379\\
f14 & 2.172 & 2.03 & 3.706 & 3.45 & 2.36 & 2.63 & 3.01 & 2.35 & 4.01 & 2.915 & 3.046\\\hline
f15 & 400 & 269.6 & 293.8 & 259 & 4.854 & 365 & 228 & 510 & 211 & 32 & 253.8\\
f16 & 93.49 & 101.6 & 177.2 & 113 & 94.76 & 143.9 & 91.3 & 95.9 & 105 & 101.2 & 109.6\\
f17 & 109 & 127 & 211.8 & 115 & 110.1 & 156.8 & 123 & 97.3 & 549 & 114.1 & 119\\
f18 & 420 & 803.3 & 901.4 & 400 & 760.7 & 483.2 & 332 & 752 & 497 & 719.4 & 439.6\\
f19 & 449 & 762.8 & 844.5 & 420 & 714.3 & 564.4 & 326 & 751 & 516 & 704.9 & 380\\
f20 & 446 & 800 & 862.9 & 460 & 822 & 651.9 & 300 & 813 & 442 & 713 & 440\\
f21 & 689.3 & 721.8 & 634.9 & 492 & 536 & 484 & 500 & 1050 & 404 & 464 & 680.1\\
f22 & 758.6 & 670.9 & 778.9 & 718 & 692.4 & 770.9 & 729 & 659 & 740 & 734.9 & 749.3\\
f23 & 638.9 & 926.7 & 834.6 & 572 & 730.3 & 640.5 & 559 & 1060 & 791 & 664.1 & 575.9\\
f24 & 200 & 224 & 313.8 & 200 & 224 & 200 & 200 & 406 & 865 & 200 & 200\\
f25 & 403.6 & 395.7 & 257.3 & 923 & 365.7 & 373 & 374 & 406 & 442 & 375.9 & 406\\\hline
\end{tabular}
\end{table*}

\section{Maximality of significance relations}

Usually in optimization we study a real-valued function $y = f(x)$ where the quest is for a value $x$ that maximizes (in case $f$ is seen as a quality or fitness function) or minimizes (in case $f$ is seen as a coast function) the function value $y$. In a number of circumstances, especially related to modern applications of computer science, this approach (then called global optimization) appears to be not fully adequate to reflect aspects of optimality like fairness, resource limitations, user preferences or simply multiple conflicting objectives. Thus, relational optimization attempts to generalize this concept of optimality by studying maximality of relations.

A (set-theoretic) binary relation between elements of a domain $A$ is a subset of $A \times A$\/, i.e. a number of ordered pairs of elements of $A$ that are considered to be in that relation with each other. This is also the way to interpret above optimality task: if $x_1$ and $x_2$ are from the domain of $f$ and we have $f(x_1) > f(x_2)$ and we are looking for maximality, then the relation $R_{opt}$ is just the real-valued larger relation between function values, and the pair $(x_1, x_2) \in R_{opt}$.

The essential point, following Suzumura's theory of social choice \cite{suzumura} is that a concept of \emph{maximality} can be considered for any relation, no matter what its domain and what the specific way of specifying related pairs. To do this, we first consider the asymmetric part $P(R)$ of a relation: it is the part of the relation where the order in the pair matters, i.e. for a relation $R$ the pair $(x,y) \in R$ but not $(y,x)$. The elements of $R$ that are not in $P(R)$ then establish the symmetric part $I(R)$ where both $(x,y)$ and $(y,x)$ belong to $R$. Thus, we have $R = P(R) \cup I(R)$ and $P(R) \cap I(R) = \emptyset$ and a convenient test for each pair in $R$ whether it belongs to the asymmetric or the symmetric part.

Now, for the asymmetric part we consider all elements of the domain that never appear in the second position: this is the \emph{maximum set} of R. If we read $(x,y) \in R$ as ``$x$ is better than/dominates/is preferred to y'' then for any element of the domain that never appears in the role of $y$ here it means that there is no better, or preferred element, or that it is non-dominated.

We see that this definition only uses the set-theoretic specification of a relation and nothing else, and includes the above example of function optimality as a special case.

It has to be distinguished from greatestness of a domain element. For a greatest element $x$ from the best set of a relation, for each $y \in A$ (including $x$ itself) we have $(x,y) \in R$ that is, using above readings, $x$ is better than any other element, preferred to any other element, or dominating any element of $A$. There are some relations between the maximum set and the best set (for example that the best set is a subset of the maximum set). In case of global optimality both basically coincide, but in the general case not. However, the specific way of defining verifiable relations in optimization problems gives preference for the concept of maximality - for the price that there is usually more than one maximal element, compared to usually empty best sets.

Now we consider maximum sets of relations within the scope of experimental algorithm analysis, taking significance into account.

\begin{definition}\label{def:1}
A \textbf{significance relation} is a family $R_{\sigma}$ of relations parameterized by the (real-valued non-negative) \textbf{significance level} $\sigma$ such that:
\begin{enumerate}
	\item For $\sigma_1 > \sigma_2$\/, $R_{\sigma_1} \subseteq R_{\sigma_2}$ and
	\item For $\sigma \rightarrow \infty$\/, $R_{\sigma} \rightarrow \emptyset$
\end{enumerate}
\noindent The relation $R_0$ is called the \textbf{base relation}.
\end{definition}

\noindent These two requirements reflect the common ideas behind significance. Take as an example the comparison of two algorithms by their average performances on a number of benchmark functions: in order to  consider algorithm $A$ really better than algorithm $B$ we set a threshold $s$ such that algorithm $A$ is only considered to be better than $B$ when having a larger average than for $B$ by margin $s$. If we increase that $s$ the number of cases where $A$ is better than $B$ will decrease --- this is the first requirement. If $s$ becomes larger, at some point, no algorithm will be better than any other by such a large margin, and the relation becomes the empty set. Last but not least, in case of $s=0$ the relation becomes the standard real-valued larger-relation that can be seen as the base relation for comparison.

Now that there is some evolution of relations over the span of confidence values, the question is what happens to their maximum sets when $\sigma$ is increasing. It needs two comments: 

\noindent (1) the definition of maximum sets applies to any relation. However, maximum sets can be empty. A sufficient condition for the existence of non-empty maximum sets for finite and non-empty domains is cycle-freeness of the relation. Cycle-freeness means that there is no sequence of one or more elements $x_i$ ($i=1, \dots, k$) where $(x_1,x_2) \in P(R), (x_2, x_3) \in P(R), \dots, (x_{k-1}, x_k) \in P(R)$ and also $(x_k, x_1) \in R$. Note that the naming ``cycle'' refers to the alternative representation of a relation as a directed graph, where cycles are equivalent to fully connected components. The definition of $P(R)$ clarifies the issue for $k=1$ and $k=2$\/, but in other cases it can become a complicated task to decide whether there are cycles or not. But even if there are cycles, it does not automatically imply empty maximum sets in specific cases. However, the empty set, or empty relation does not contain cycles for sure, and therefore by a continuity argument, for any significance relation there is some $\sigma$ such that for this value and all larger the relation will be cycle-free and the existence of maximum elements is guaranteed.

\noindent (2) On the other hand, for the empty relation each element of the domain is maximal. Under the additional constraint that $R_{\sigma}$ for each $\sigma$ is an asymmetric relation, then the size of maximum sets increases monotonically with increasing $\sigma$.

Taking both arguments together, we can identify a significance level where the relation becomes cycle-free (or one where the maximum set is non-empty) while the size of the maximum set is at the lowest non-negative level. From this we can identify the ``best algorithms'' by that maximum set.

At the end, we have to consider how to design an appropriate relation. This will be demonstrated in the next section, where we are going to apply the theoretical foundation given in this section to the CEC'05 benchmark.

\begin{figure*}[htb]
\centering
\includegraphics[width=12cm]{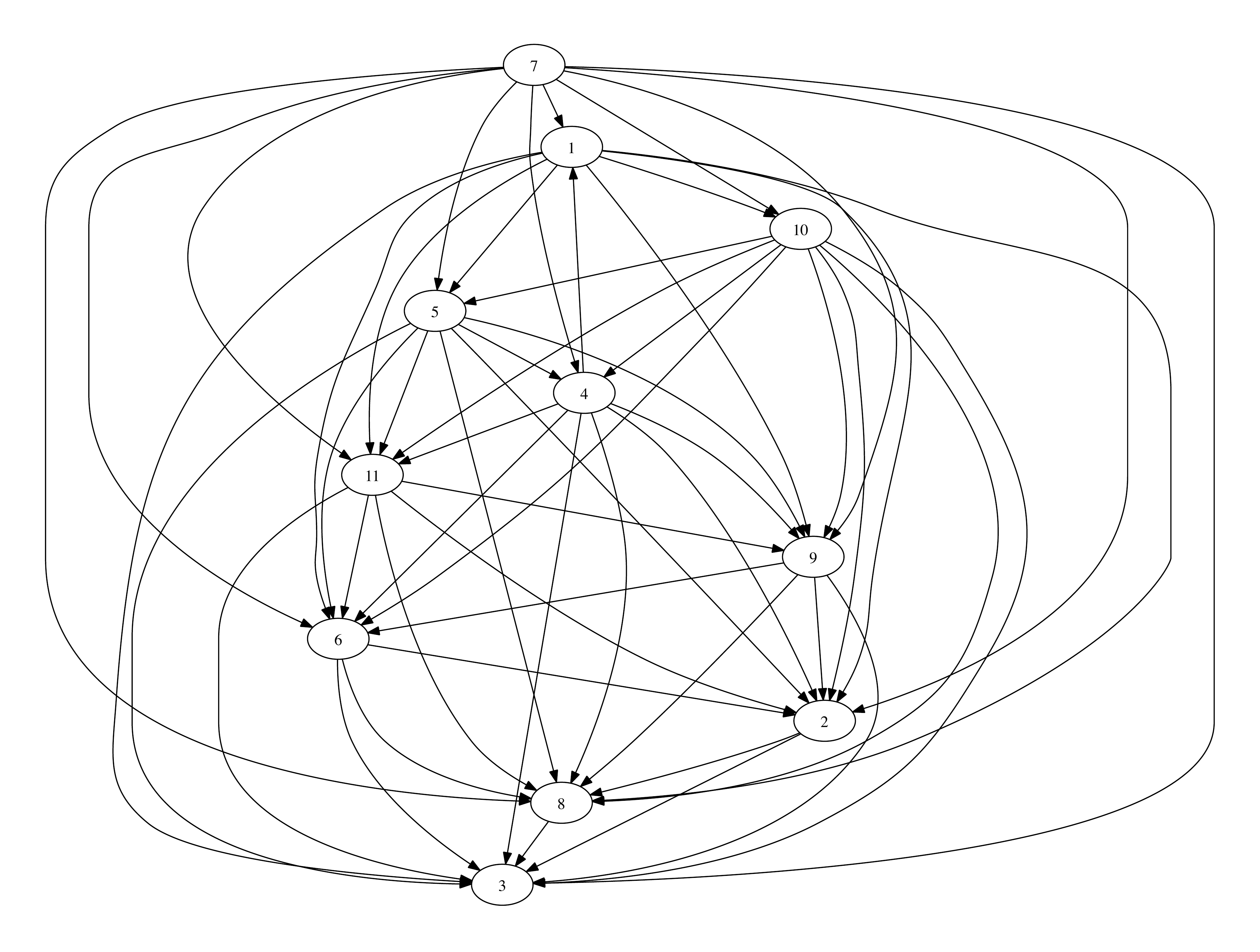}
\caption{Graph of the significance relation for CEC'05 benchmark functions and test algorithms and significance level 10.}
\label{fig:th10}
\end{figure*}

\section{Application to CEC'05 benchmark}

The CEC'05 benchmark is composed of 25 functions that fall into three categories: functions 1 to 5 are unimodal functions, functions 6 to 14 multi-modal functions and the remaining functions 15 to 25 are so-called hybrid composition functions. For details see the corresponding technical report \cite{suganthan2005problem}. For the contest, 11 algorithms were selected. Achieved performance values for dimension 10 versions of all 25 problems can be seen in Table 1 (these are the same values that were studied in \cite{garcia2009study}). A quick glance on this table already reveals a number of issues that make sure that the evaluation of the outcome of the experiment is not straightforward. At first, the performances differ largely in order of magnitude. Values $1 \times 10^{-9}$ represent cases were the problem was ``fully solved'' i.e. the algorithm were stopped at this point. Such values are predominant in the first category of unimodal functions. For functions of the second category we find varying performance, while the third category provides most challenging functions, and algorithms seems to yield comparable bad performance values.

If considering a base function for the specification of a significance relation, this makes clear that the average of these values per algorithm is not suitable due to the different scale of the performance values for the particular functions, as well as any reference to Euclidian distances between points with performance values as components. In fact, what we need is

\begin{itemize}
	\item Horizontal scale-freeness: each problem function provides its own scale for the typical range of performance values. It can be achieved by considering the relative gains of performance instead of taking reference to absolute values.
	\item Vertical scale-freeness: the evaluation also needs to take the differing performance scales of different problems into account. This can be achieved by awarding rank points, as it is done in the Borda count method for voting (see \cite{borgers2009socialchoice} for a gentle introduction into this topic), or the Wilcoxon signed rank test for statistical significance.
\end{itemize}

Based on these arguments, the proposed significance relations for given significance level $\sigma$ applied to two vectors $x$ and $y$ both of dimension $n$ and with positive components is as follows:

\begin{enumerate}
	\item Compute the ranking vector $r$ where $r_i = \max[x_i/y_i, y_i/x_i]$ ($i=1,\dots,n$) as well as the signature vector $s$ where $s_i$ is +1 if $x_i > y_i$\/, -1 if $x_i < y_i$ and 0 if $x_i = y_i$.
	\item Sort the components of $r_i$ in non-decreasing order to yield the vector $\tilde{r}_{(i)}$.
	\item The award points vector $a$ is the vector $(1, 2, \dots, n)$ permuted in the same way as the permutation leading from $r$ to $\tilde{r}$.
	\item Compute the scalar product $D = a \cdot \tilde{s}$ (where $\tilde{s}$ is the sorted version of $s$). In case of ties, collect the corr. award points and share equally between $x$ and $y$.
	\item If $D \ge \sigma$ then $x$ is in significance relation to $y$\/, otherwise not.
\end{enumerate}

In summary, the ratios between corr. components of $x$ and $y$ (the larger one divided by the smaller one) are sorted by size, and signed award points are given two $x$ or $y$\/, 1 for the smallest ratio, 2 for the second smallest etc. If $x$ had the larger component, the sign is +1, -1 otherwise. For example, consider the vectors $x = (2, 8, 6)$ and $y = (4, 2, 3)$ and significance $\sigma = 2$. Then $r = (4/2, 8/2, 6/3) = (2, 4, 2)$ which is sorted as $\tilde{r} = (2, 2, 4)$ and the corr. award vector is $s = (1, 3, 2)$ ($(2, 3, 1)$ is also possible but would not affect the result). The sign vector registers which vector had the larger component: $s = (-1, 1, 1), \tilde{s} = (-1, 1, 1)$. Since the two first components of $\tilde{r}$ are equal, the total award points $1+2 = 3$ are shared equally as $3/2 = 1.5$ and $x$ receives $1.5$ (for its second component) and 3 (for the third component) while $y$ receives $1.5$ for the second component. The total is $D = 1.5 + 3 - 1.5 = 3$. Since 3 is larger than the confidence, $x$ is in relation to $y$ (or: $x$ is better than $y$ with significance at least 2).

\begin{figure*}[htb]
\centering
\includegraphics[width=12cm]{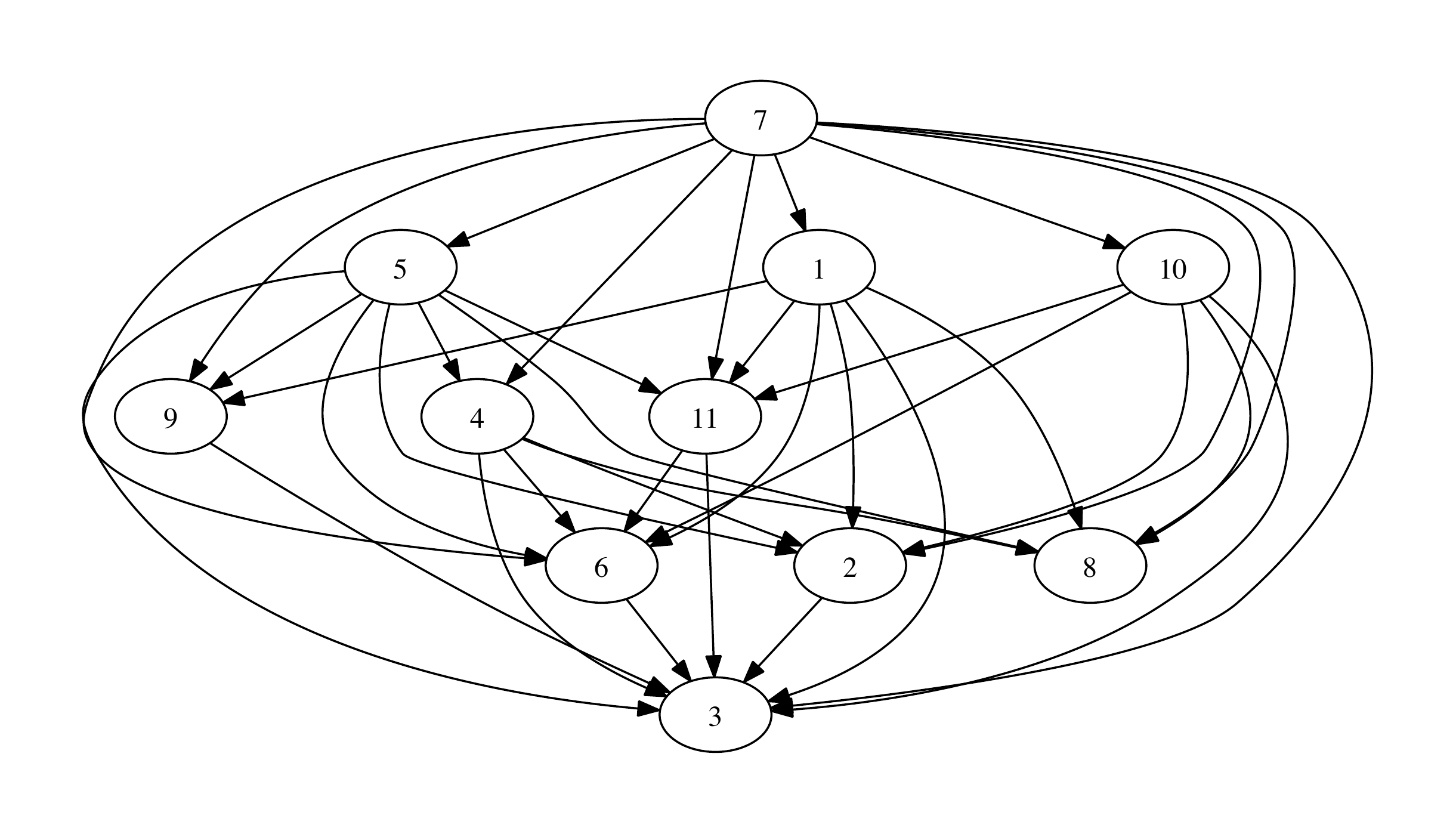}
\caption{Graph of the significance relation for CEC'05 benchmark functions and test algorithms and significance level 60.}
\label{fig:th60}
\end{figure*}

We apply this relation to the CEC'05 benchmark function results for dimension 10 on the 11 test algorithms. It means that we select various increasing values of $\sigma$ until the relation becomes cycle-free, and then look for the maximal (or: non-dominated) elements. Figure \ref{fig:th10} shows the result for $\sigma = 10$. It is shown as a directed graph where a directed edge (or arc) refers to the relation between the corr. algorithm to exist (the algorithms are given as numbers in the order of the header row of Table 1). A number of things can be seen:

\begin{itemize}
	\item Algorithm 7 is a clear winner, as it is in significance relation to all other algorithms.
	\item Algorithm 3 is a clear looser, since it is dominated by all other algorithms.
	\item There is one cycle in the graph, linking algorithms 1, 10, 5, 4 and back to 1.
	\item Algorithms 2, 6, 8, 9 and 11 appear to be on lower rank than this cycle, but of higher rank than algorithm 3.
\end{itemize}

If we increase $\sigma$ at around 15 the relation becomes cycle-free and above four ranks consolidate (not shown here for space reasons). For $\sigma = 60$ (Fig. \ref{fig:th60}) we can distinguish five ranks (7 as top performer as before, next rank 5, 1 and 10, next rank 4, 9, 11, then 2, 6 and 8, and last as well as least 3). Note that algorithm 4 was part of the cycle before, now it is ranked below all other members of the cycle. For $\sigma = 100$ it will shrink to 4 ranks and at some point all but algorithms 7 and 3 will be on the same second rank.

\section{Discussion}

The analysis of the CEC'05 example shows that the procedure is feasible and allows for a ranking of a number of algorithms. Here some related thoughts and considerations:

\noindent (1) The procedure possibly assigns same rank to several algorithms. In the worst case, all algorithms or its larger part can be on the same rank - that is when for the smallest significance level $\sigma$ where the relation becomes cycle-free suddenly many elements are not dominated by any other element. While this is possible in theory, in this case the design of the benchmark might be questioned since it also indicates a lack of distinctiveness from the selected benchmark functions (for example, when all algorithms can solve all problems). As the result for the CEC'05 benchmark shows, there can be even a best algorithm.

\noindent (2) The computational effort of the evaluation is rather small - as long as the number of algorithms is bounded. It needs pairwise comparison and a linear comparison procedure. However, even polynomial complexity can become hard when the number of algorithms is of order 1 Million or so, since then there would be $10^{12}$ comparisons, an effort far beyond the available computational power these days. In this case, meta-heuristic algorithms can be designed to approximate maximum sets of general relations \cite{koppen2011meta}.

\noindent (3) The selection of $\sigma$ depends on the specific choice of algorithms. It means if the set of algorithm changes, the selection procedure for $\sigma$ can also give a different value. So questions like this may come up: algorithm 7 was the winner of the CEC'05 benchmark. What to do to make an even better algorithm? The answer depends much on the benchmark functions themselves, but a direct answer is: either to be better than algorithm 7 for all functions, or being better for functions where algorithm 7 (compared e.g. to next rank algorithms) is weak.

\section{Summary}

A method for the evaluation of benchmark results for function optimization has been presented. It is based on the general concept of significance relations: two algorithms are in such a relation if one performs better than the other by a gain that is numerically represented as significance level. When the significance level is increased there must be a level where the relation becomes cycle-free and provides a full ranking of all algorithms (thus also providing a partial order of the algorithms). The procedure distinguishes from the classical ``compare the average performances'' in the sense that it accounts for significance while switching to a relational mode (instead of absolute mode) of comparison. The feasibility of the approach has been shown by using test results for the CEC'05 benchmark.

% \section{Conclusion}
% The conclusion goes here. this is more of the conclusion
% 
% % conference papers do not normally have an appendix
% 
% 
% % use section* for acknowledgement
% \section*{Acknowledgment}
% more thanks here



%\bibliographystyle{IEEEtran}
% \end{thebibliography}

\bibliographystyle{IEEEtran}
\bibliography{algeval}

\end{document}